# Properties of the angular gap in a one dimensional photonic band gap structure containing single negative materials.

**Munazza Zulfiqar Ali and Tariq Abdullah.**
Centre of Excellence in Solid State Physics, Quaid-i-Azam Campus,
Punjab University, Lahore, 54590, Pakistan.

## Abstract.

The linear properties of the angular gap in a one-dimensional photonic band gap structure containing single negative material layers are investigated. This gap forms at oblique incidence due to total internal reflection into the air when the Snell's law breaks down and its lower edge occurs at the frequency where the refractive index of one or both layers of the structure approaches zero. This gap is found to be highly sensitive to the incident angle and the polarization of light, but is not affected by thickness ratio of the layers. It is also shown that the electric field is extremely enhanced at the low frequency edge of this gap for TM polarization. This highly enhanced electric field can be used for certain applications.

**PACS:** 42.70.Qs, 77.84.Lf, 78.20.Ci

**Keywords:** Photonic band gap structures, single negative materials, angular gap.

## 1. Introduction.

The experimental realization of double negative (DNG) and single negative (SNG) metamaterials and their inclusion in photonic band gap (PBG) structures has given rise to mechanisms other than the conventional Bragg reflections to produce photonic band gaps [1-7]. The theoretical models which describe the electric permittivity and magnetic permeability in DNG and SNG metamaterials are essentially dispersive. A zero-n gap [8,9] is found in the photonic band gap structure containing alternate DPS (double positive i.e. regular materials) and DNG layers in that frequency range in which the average refractive index of the structure becomes zero. A zero-$\phi_{eff}$ [10,11] gap is found in the structure containing alternate ENG (epsilon- negative material i.e. in which electric permittivity is negative but magnetic permeability is positive) layer and MNG (mu- negative material i.e. in which magnetic permeability is negative but electric permittivity is positive) layers. It results due to a mismatch in the local phase shifts of the two layers around the wave impendence matching frequency. It has been found that these gaps are relatively insensitive to polarization of light, incident angle, disorder and rescaling [5-11]. Recently [12] the properties of a new type of gap associated with DNG materials have been explored. It was found that a new gap forms at oblique incidence in a one-dimensional DNG-DPS photonic band gap structure in the frequency region where the refractive index of the negative index material becomes approximately zero. It forms due to total internal reflection into the air when the Snell's law breaks down. This angular gap has very narrow band edge resonances and was found to be highly sensitive to the incident angle and the polarization of the incident light. Further it was seen that the electric field enhances to a very high value inside the structure at the edges of this gap for transverse magnetic (TM) incident light. In that work the authors suggested exploiting these properties for second harmonic generation. In the present work we explore the properties of a gap, which is found at oblique incidence in a photonic band gap structure composed of alternate ENG and MNG layers. The physics involved in the formation of this gap is same as that in the case of Ref.[12] but its properties differ in some respects as the structure under consideration has different characteristics. The existence of such a gap was pointed out in Ref.[13] but its properties has not been explored to the best of our knowledge. We have found that the low frequency edge of this gap remains relatively fixed as the angle of incidence varies and the useful property of extra ordinary high electric field localization can be achieved at the low frequency edge of this angular gap for TM polarization which is even higher as compared to that in the case of Ref.[12]. The motivation behind this study is the fact that the fabrication techniques of ENG or MNG materials are less intricate than those to produce a DNG material. So a gap in ENG-MNG photonic band gap structure with similar properties is more useful than having such gap in a DPS-DNG structure. Also a periodic structure consisting of ENG and MNG layers can show a restricted equivalence with a homogenous slab of DNG material as shown in some recent studies [14,15]. More over the presence of high electric field localization at the low frequency edge of this gap can remarkably decrease the thresholds of input intensity for gap soliton [16,17] formation in the case of nonlinear wave propagation and the associated phenomenon of optical bistability can take place at extremely low values of input intensity which is a highly desirable situation for the process of optical switching. However we have considerer only linear wave propagation in this paper. The nonlinear wave propagation will be considered in subsequent work.

## 2. **Mathematical formalism.**

We consider a one-dimensional structure of alternate layers A and B. We have initially represented the values of electric permittivity and magnetic permeability of the layers by a lossless Drude model to concentrate on the emergence of new features; however the effect of losses is discussed later in this work.

$$\varepsilon_A = 1 - \frac{\omega_{eA}^2}{\omega^2}, \qquad \mu_A = \alpha \qquad (1)$$

$$\varepsilon_B = \beta, \qquad \mu_B = 1 - \frac{\omega_{mB}^2}{\omega^2} \qquad (2)$$

Here we consider the propagation of a plane wave incident at an angle $\theta$ with respect to the normal to the surface of the structure. The tangential (to the interfaces) components of the electric and magnetic fields across the jth layer of width $d_j$ are related by the following transfer matrix [13,18].

$$m_j = \begin{bmatrix} \cos k_j d_j & -\frac{1}{q_j} \sin k_j d_j \\ q_j \sin k_j d_j & \cos k_j d_j \end{bmatrix} \qquad (3)$$

Where

$$k_j = \frac{\omega}{c}\sqrt{\mu_j \varepsilon_j}\sqrt{1 - \frac{\sin^2\theta}{\mu_j \varepsilon_j}} \qquad (4)$$

And

$$q_j = \frac{\sqrt{\varepsilon_j}}{\sqrt{\mu_j}}\sqrt{1 - \frac{\sin^2\theta}{\mu_j \varepsilon_j}} \quad \text{for the TE field} \qquad (5)$$

$$q_j = \frac{\sqrt{\mu_j}}{\sqrt{\varepsilon_j}}\sqrt{1 - \frac{\sin^2\theta}{\mu_j \varepsilon_j}} \quad \text{for the TM field} \qquad (6)$$

The tangential components of electric and magnetic field at the incident side $z = 0$ and at the transmitted side $z = L$ are related by the following matrix equation:

$$\begin{bmatrix} E_1 \\ H_1 \end{bmatrix}_{z=0} = M \begin{bmatrix} E_N \\ H_N \end{bmatrix}_{z=L} \qquad (7)$$

Where

$$M = \prod_{j=1}^{N} m_j \qquad (8)$$

Here $N$ is the total number of layers in the structure. The transmission coefficient $T$ of the finite structure is calculated by applying the boundary conditions at the incident and the transmitted ends and is given by the following expression:

$$T = \frac{2q_0}{[(q_0 M_{11} + q_0 M_{22}) - (q_0^2 M_{12} + M_{21})]} \qquad (9)$$

Where $q_0 = \sqrt{1 - \frac{\sin^2\theta}{\varepsilon_0 \mu_0}} = \cos\theta$, as there is air i.e. $\varepsilon_0 = \mu_0 = 1$ on the incident and the transmitted side of the structure and $M_{ij}$ are the elements of the matrix $M$.

### 3. **Results and discussions.**

Initially we have considered the case when the electric plasma frequency ($\omega_{eA}$) in the ENG layer and the magnetic plasma frequency ($\omega_{mB}$) in the MNG layer have same values. The values of the magnetic permeability $\alpha$ in ENG layer and electric permittivity $\beta$ in the MNG layer are also taken to be the same. This means that the refractive index of two layers is same over all frequency ranges. In our computational work we have used the dimensionless units i.e $W = \frac{\omega d}{c}$ and $D_j = \frac{d_j}{d}$, $j = A, B$, where $d = d_A + d_B$ and $c$ is the velocity of light. The advantage of using these dimensionless units is that the widths belonging to any length scale and the corresponding values of frequencies can fit these calculations; however we want to mention that since the SNG metamaterials have been experimentally realized in GHz frequency rang, the following set of realistic values can fit these calculations:

$$\omega_{eA} = \omega_{mB} = 2\pi \times 3.34 \text{ GHz}, \quad \alpha = \beta = 1, \quad d_A = 4 \text{ mm}, \quad d_B = 6 \text{ mm} \quad (10)$$

Fig.1a shows the case for a normally incident wave ($\theta = 0°$) for a structure consisting of 10 periodic units ($N$=20). A zero-$\phi_{eff}$ gap appears in the frequency region where the structure is composed of SNG layers and a Bragg gap appears in the region where the refractive index of both layers is positive. Fig.1b shows the case for oblique incidence for TE polarization at $\theta = 30°$, a new gap appears whose lower edge is located at the frequency which is nearly equal to plasma frequencies in the two SNG layers. Fig.2a shows the angular dependence of this gap. The high frequency edge shifts higher on the frequency axis as the angle of incidence increases. This figure (Fig.2a) also compares the angular dependence of this new angular gap with that of the zero-$\phi_{eff}$ gap which is found to be insensitive to the incident angle. The formation of this angular gap can be explained in a simple manner. In a bulk ENG or MNG metamaterial evanescent modes are present. The appearance of the propagating modes in the ENG-MNG multilayered structure can be understood in terms of the tight binding model of solid state physics according to which the propagating and the decaying modes resemble each other over small range of distance, of the order of the width of the layers in the structure. For frequencies below the low frequency edge of this angular gap the layers are SNG whereas above the high frequency edge, the refractive index of the layers is positive. For frequencies surrounding this gap the absolute value of their refractive index is less than unity i.e $|\sqrt{\varepsilon_j \mu_j}| < 1$. As the surrounding medium is air, the light is entering from a medium of high refractive index to a medium of low refractive index. When the refractive index of both layer of the structure becomes nearly zero at oblique incidence, the Snell's law breaks down i.e. for $\sin\theta/n_j = \sin\theta_j/1$ ($j$=A,B and $\theta \neq 0$) no physical solution for $\theta_j$ exists when $n_j$ approaches zero, so the phenomena of total internal reflection into the air takes place. At this particular frequency, due to the photonic band gap effect of the periodic structure a gap is opened up for a certain range of frequencies. This gap has a true mirror like behavior i.e. its transmissivity can be shown to be nearly zero which is not affected by the number of layers of the structure, however adding more layers will increase the spectral width of the gap. It should be remembered that the structure is not showing metallic behavior [19] in the frequency range in which this angular gap exist otherwise the gap would have existed for normal incidence as well. Interestingly this angular gap retains its existence at oblique incidence when the zero-$\phi_{eff}$ gap ceases to exist i.e. when the impedance and the local phase shifts in the two layers are in perfect match (i.e when the widths of the two layers are equal in this particular case) as shown in Fig.2b (dotted lines). This figure also shows that this angular gap is insensitive to the ratio of the widths of the two layers whereas the zero-$\phi_{eff}$ gap widens as this ratio increases.

Figure 3a shows the case when the incident light is TE (continuous) and when it is TM (dotted). In this particular case when the refractive indexes of the two layers are same, this angular gap is found to be insensitive to the polarization to the incident light. But if the refractive index of the two layers are not same (either the values of electric plasma frequency in the ENG layer and the magnetic plasma frequency in the MNG layer are different or the values of magnetic permeability in ENG layer and electric permittivity in the MNG layer are unequal) then the gap shows its sensitiveness to the incident polarization as shown in Fig.3b. The detail of the parameters is given in the figure caption. For TE mode the lower edge of the angular gap lies at the point when the incident frequency approaches the magnetic plasma frequency of the MNG layer, whereas for TM mode the lower edge of the angular gap lies at the point when the incident frequency approaches the electric plasma frequency of the ENG layer.

There are a few exceptionally narrow band edge resonances at the lower edge of this angular gap as shown in Fig. 4. These resonances are separated by frequencies of the order of $10^{-3}$ (in dimensionless units) on the frequency axis and their positions are sensitive to the angle of incidence. We have plotted the absolute value of electric field (in units of the incident electric field) inside the structure at the lower edge of the zero-$\phi_{\text{eff}}$ gap (Fig.5a) and at the lower edge of the angular gap for TE mode (Fig.5b) and for TM mode (Figs.5c,5d). This shows that the electric field localizes and is enhanced to an extremely high value at the lower edge of the angular gap for TM polarization. The reason for this high electric field enhancement at the low frequency edge can be understood in terms of the boundary condition of orthogonal component of electric field for TM polarizations which can be written as:

$$\varepsilon_A E_{\perp A} = \varepsilon_B E_{\perp B} \qquad (11)$$

It means that the orthogonal component of the electric field in layer A is $\varepsilon_B/\varepsilon_A$ times that in layer B. As the lower edge of this angular gap occurs at the point where $\varepsilon_A$ approaches zero, so the ratio $\varepsilon_B/\varepsilon_A$ has a very high value(nearly 100)at this point. This makes the electric field discontinuous across the interface localizing it in layer A. As the structure is periodic, the multiple reflections also contribute towards the field enhancement and force the field into bell-shaped envelope across the structure as shown in Figs. 5(a-d) the The magnitude of the electric field is enhanced by a factor of approximately 400 in Fig.5c and by a factor of 175 in Fig. 5d( which is plotted for a structure containing only 5 periodic units). This high electric field localization can be utilized for many practical applications. It can greatly enhance the nonlinear effects which can be further exploited for the purpose of second harmonic generation under phase matching conditions and also to reduce the threshold of optical bistability. In the case of Ref.[12], the electric field is enhanced at both ends of the angular gap as the frequency at which the refractive index of the negative index material layer becomes zero is located in the middle of the gap whereas in the structure considered here the frequency at which the refractive index of the layers become zero occurs at the low frequency end of the gap so the electric field is extremely enhanced at this end but the enhancement of the electric field does not take place at the high frequency edge of the angular gap as the values of electric permittivities and magnetic permeabilities of both layers are comparable at this point.

Uptill now we have considered a highly idealized situation in which the role of losses has been neglected. Finally we want to make a few comments on the effect of losses i.e. the electric permittivity of layer A and magnetic permeability of the layer B are represented by a lossy Drude model:

$$\varepsilon_A = 1 - \frac{\omega_{eA}^2}{\omega(\omega + i\gamma_A)} \quad , \quad \mu_B = 1 - \frac{\omega_{mB}^2}{\omega(\omega + i\gamma_B)} \qquad (12)$$

Figs.6 (a-c) compare the effect of losses on the lower gap edge resonances for TM polarization when (a) $\gamma_A = \gamma_B = 0$, (b) $\gamma_A = \gamma_B = 2\pi \times 1\,\text{MHz}$ and (c) $\gamma_A = \gamma_B = 2\pi \times 10\,\text{MHz}$. The first figure (from left) is plotted for $N=20$. In this case the first few resonances disappear as the losses increases, however the higher resonances compare well with the ideal situation. So the enhancement of the electric field of the order of that shown in Fig.4 (c) can not be achieved in real structures as the presence of losses destroy the first few resonance peaks located at the lower edge, however if we decrease the length of the periodic structure, the effect of losses decreases appreciably. The second figure has been plotted for a structure composed of only 5 periodic units ($N=10$). Approximately 80 percent of the transmission is retained at the first resonance peak when a loss term of $\gamma_A = \gamma_B = 2\pi \times 10\,\text{MHz}$ is present. The electric field enhancement factor in a structure composed of only 5 periodic units is nearly 175, which is a remarkable figure. So periodic structures consisting of only few periodic units can survive losses and yet provide a high value of electric field enhancement at the lower edge of the angular gap.

**Conclusions.**

In summary, the properties of a new gap, which is found at oblique incidence in a one-dimensional photonic band gap structure containing SNG materials, are investigated. This angular gap originates due to total internal reflection into the surrounding air and its lower edge lies at the frequency at which the refractive index of one or both layers of the structure approaches zero. This gap is highly sensitive to the incident angle and polarization of the incident light, however it does not show its dependence on the ratio of the thicknesses of the two layers. The electric field localizes to extremely high values at the low frequency edge of this angular gap for TM polarizations in the ideal situation of no losses. The presence of losses destroys the resonance peaks at the edge of the gap. However a periodic structures consisting of only few periodic units can survive losses and yet provide a high value of electric field enhancement at the lower edge of the angular gap. It is suggested that this extremely high electric field localization can cause a tremendous reduction in the threshold of input intensity for the gap soliton formation process in the case of nonlinear wave propagation. So the associated phenomenon of optical bistability can take place at extremely low values of input intensity which is a highly desirable situation for optical switching. Further investigations along these lines are under consideration.

**References.**

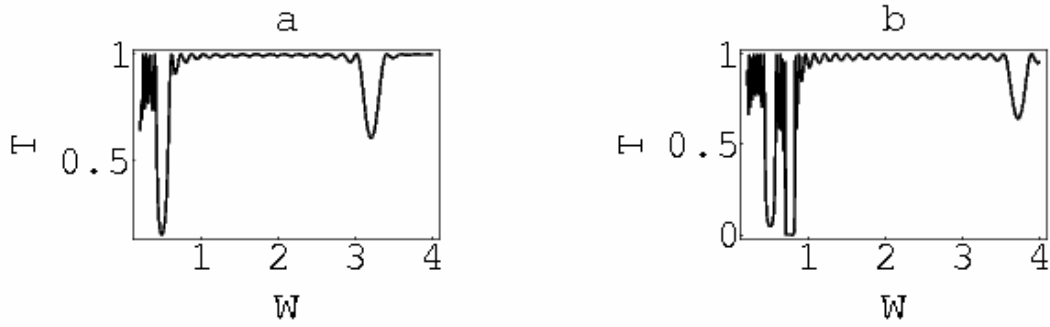

Fig.1 The transmission spectrum of the structure for $W_{eA} = W_{mB} = 0.7$, $D_A = 0.4$, $D_B = 0.6$ (in dimensionless units given in the main text) for TE polarization for N=20(a) the case for a normally incident wave (b) for oblique incidence at $\theta = 30°$.

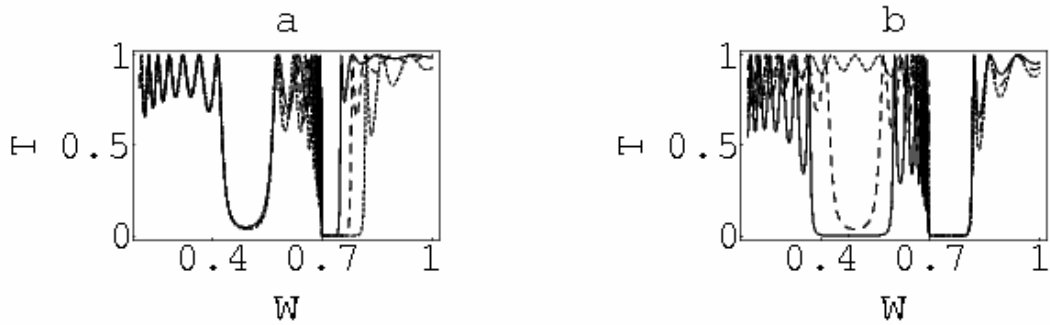

Fig.2 (a) The angular dependence of the angular gap where the continuous, dashed and dotted lines correspond to $\theta = 20°, 30°, 40°$ respectively. (b) The dependence of the angular gap on the ratio of the thickness of the two layers where the continuous, dashed and the dotted lines correspond to $D_A : D_B$ taken to be 3:7, 4:6, 5:5 respectively, other parameters are same as in Fig.1.

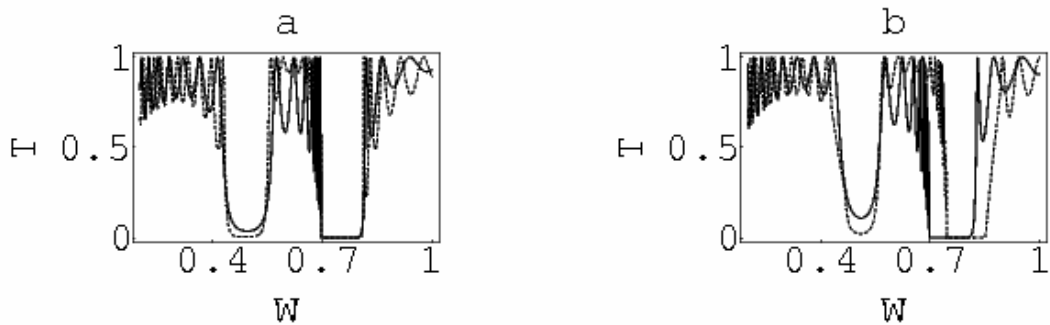

Fig.3 (a) The transmission spectrum for the parameters given in Fig.1 at $\theta = 30°$ for TE polarization (continuous curve), TM polarization(dotted curve) In (b) $W_{eA} = 0.75$, N=20 for both cases.

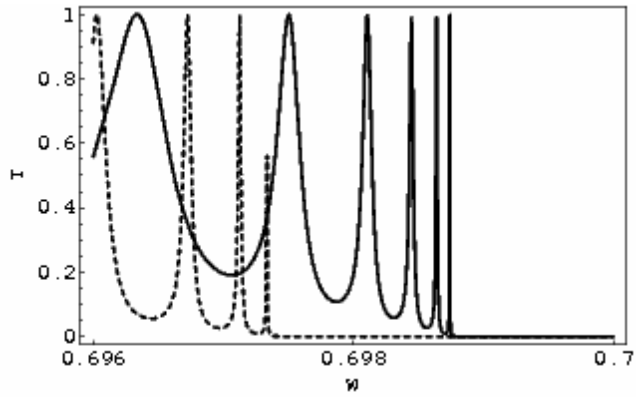

Fig.4 The figure shows the narrow resonance positions at the low frequency gap edge of the angular gap where continuous curve represent the case when $\theta = 20°$ and the dotted lines represent the case when $\theta = 30°$, other parameters are same as in Fig.1

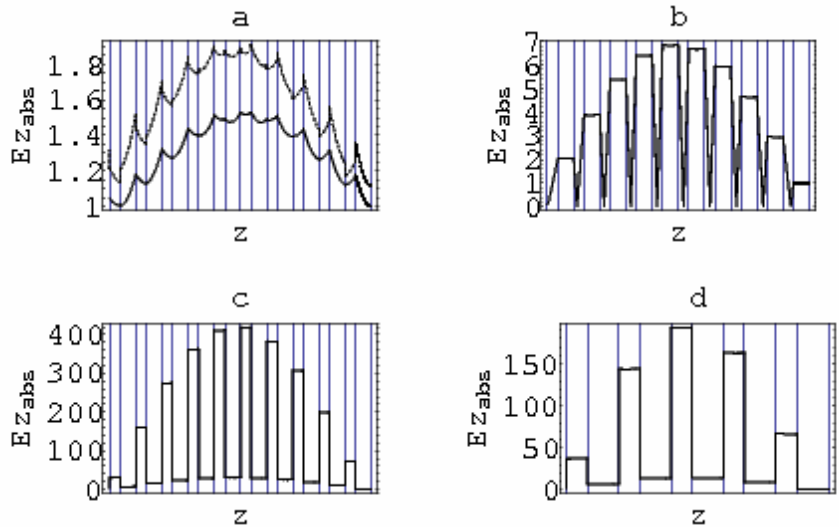

Fig.5. The figure shows the field localization at the low frequency edge (first resonance position ) of (a) zero-$\phi_{eff}$ gap where the continuous curve corresponds to TE mode and the dotted curve represents the TM polarization (b) angular gap at $\theta = 30°$ for TE mode (c,d) angular gap for TM mode, $N=20$ for Figs.(a,b,c) and $N=10$ for Fig. (d), on the vertical axes the absolute value of electric field is normalized with respect to absolute value of the incident field. Thinner layers are A layers and thicker layers are B layers. Other parameters are same as in Fig.1

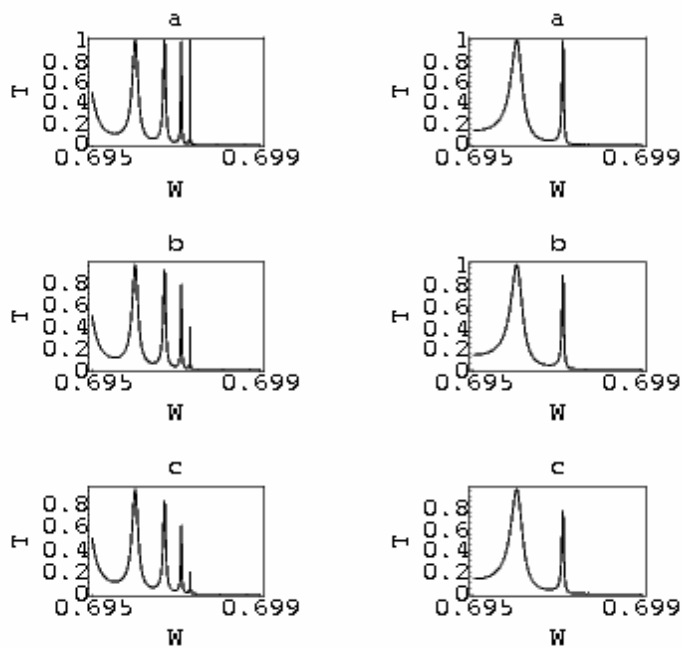

Fig.6. The effect of losses on the low frequency gap edge resonances of the angular gap for the structure consisting of $N=20$ (left Fig.) and $N=10$ (right Fig.) where (a) corresponds to $\gamma_A = \gamma_B = 0$ (b) corresponds to $\gamma_A = \gamma_B \approx 2\times 10^{-4}$ (c) corresponds to $\gamma \approx 2\times 10^{-3}$ (in dimensionless units). All other parameters are same as in Fig.1